# Sub-Doppler Double-Resonance Spectroscopy of Methane Using a Frequency Comb Probe

Aleksandra Foltynowicz[1], Lucile Rutkowski[2], Isak Silander[1], Alexandra C. Johansson[1], Vinicius Silva de Oliveira[1], Ove Axner[1], Grzegorz Soboń[3], Tadeusz Martynkien[4], Paweł Mergo[5], Kevin K. Lehmann[6]

[1]*Department of Physics, Umeå University, 901 87 Umeå, Sweden*
[2]*Univ Rennes, CNRS, IPR (Institut de Physique de Rennes)-UMR 6251, F-35000 Rennes, France*
[3]*Laser & Fiber Electronics Group, Faculty of Electronics, Wrocław University of Science and Technology, 50-370 Wrocław, Poland*
[4]*Faculty of Fundamental Problems of Technology, Wrocław University of Science and Technology, 50-370 Wrocław, Poland*
[5]*Laboratory of Optical Fiber Technology, Maria Curie-Sklodowska University, pl. M. Curie-Skłodowskiej 3, 20-031 Lublin, Poland*
[6]*Departments of Chemistry & Physics, University of Virginia, Charlottesville, VA 22904, USA*
*e-mail address: aleksandra.foltynowicz@umu.se; kl6c@virginia.edu*

We report the measurement of sub-Doppler double-resonance transitions in methane using a 3.3 μm continuous wave optical parametric oscillator to pump transitions in the fundamental $v_3$ band and a 1.67 μm frequency comb to probe ladder-type transitions in the $3v_3$ band over 200 cm$^{-1}$ of bandwidth. We detected 36 ladder-type resonances for 9 pumped transitions with average center frequency accuracy of 1.7 MHz, limited by the pump frequency stability. The lines are assigned using the TheoReTS line list, and the intensity ratios for different relative pump/probe polarizations. This method provides accurate assignment of highly excited energy levels that cannot be done using high temperature spectra.

Methane has long been a molecule of intense scientific interest [1,2]. It is the lightest stable spherical top molecule and it is used as representative of molecules with tetrahedral symmetry. It is an increasingly important fossil fuel and green house contributor [3]. It is produced and consumed by many anaerobic organisms [4]. It is a dominant source of the opacity of many of the planets in our solar system [5] and has been detected in hot-Jupiter exoplanets [6-8]. Despite this, the energy level structure of methane is far from fully understood and the modelling of observed high-temperature spectra is limited by the lack of reference laboratory data. Theoretical spectra can be produced using the parameters from the Exomol [9] or TheoReTS (Theoretical Reims-Tomsk Spectral Data) [10,11] line lists, but firmly assigned experimental transitions belonging to hot bands are needed to judge the accuracy of these predictions. However, warming up the sample to the relevant temperatures in controlled conditions is an experimental challenge and produces spectra with a high transition density that are difficult to assign [12-14].

The dominant spectroscopic signature of methane is its vibration-rotation spectrum, including weak overtone bands extending into the visible spectral region. Methane has a series of Fermi, Darling-Dennison and Coriolis resonances that cause strong mixing of normal mode states leading to clumps of states, known as polyads [15]. Rotationally resolved spectra have been reported up to 16,180 cm$^{-1}$ [16]. At present, only rovibrational spectra up to the tetradecad polyad (<6200 cm$^{-1}$) are satisfactorily assigned [17]. The complexity of CH$_4$ spectra and their analysis rises rapidly with increasing excitation.

Double resonance (DR) spectroscopy [18] is a powerful tool for assignment of highly perturbed spectra, even those produced by chaotic dynamics where the patterns that underlie traditional assignment methods are lost [19]. It provides a way to use an already assigned transition to unambiguously identify the lower state quantum numbers of measured spectra. Ideally, the population of a single quantum state is depleted and another increased by a "pump" transition, and this population change is used to modify the strength of other transitions observed using a "probe". The already assigned transition can be used either as pump or probe, and Fig. 1(a) shows the case when the pump addresses assigned transitions in the fundamental C-H stretching band. Two types of DR excitations can be observed: V-type, in which the pump and probe transitions share a common depleted rotational state of the ground vibrational states; and ladder-type, in which the upper level of the pump transition is the lower level for the probe. When a monochromatic pump is used, only a narrow velocity group of molecules is excited, and the resulting probe transitions are Doppler-free, with a width limited by the homogeneous broadening [20]. This is particularly advantageous for CH$_4$, where many transitions overlap in the Doppler-broadened spectra. Moreover, the ratio of the probe transition intensities measured with different relative polarizations of the two fields depends upon the change in rotational quanta of the pump and probe transitions [21] and can be used to assign the quantum number of the upper state.

In V-type excitations, the probe transitions appear in linear absorption spectra as dips in the Doppler-broadened lines. The ladder-type excitations, on the other hand, allow observation of sub-Doppler transitions to levels that are not allowed from the ground vibrational state. In the case of $CH_4$, ground state transitions are allowed only to vibrational states with $F_2$ symmetry, but from the $F_2$ symmetry $\nu_3$ level transitions are allowed to vibrational states with $A_1$, $E$, $F_1$ and $F_2$ vibrational symmetries (but not $A_2$).

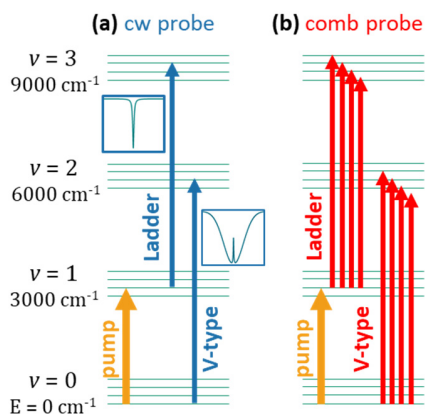

FIG. 1. (Color online) The energy level structure of the $\nu_3$ stretching bands of methane with pump (orange) and probe transitions indicated for (a) a single frequency probe (blue) and (b) a frequency comb probe (red).

Double-resonance spectroscopy of $CH_4$ was previously reported by De Martino et al. in the 1980's [22]. This pioneering work utilized nanosecond pump and probe pulses in the 1.6 μm and 3.3 μm spectral regions, respectively. Both the sensitivity (few per cent absorption noise) and the spectral resolution (>0.12 cm$^{-1}$) were limited by the optical sources. Such resolution is substantially worse than the Doppler width of methane transitions (~0.01 cm$^{-1}$ full width at 3.3 μm and 300 K) as well as the line spacing in most of the observed $CH_4$ spectra.

Here we use a high-power 3.3 μm continuous wave pump and a 1.67 μm frequency comb probe [see Fig. 1(b)] to record sub-Doppler DR transitions of methane over 200 cm$^{-1}$ of bandwidth with 1.7 MHz (5.6×10$^{-5}$ cm$^{-1}$) center frequency accuracy. The ladder-type probe transitions reach ro-vibrational states in the ~9000 cm$^{-1}$ energy region and most of the observed transitions were previously unassigned, with the exception of those identified by the previous DR experiments. The frequencies and intensities of the ladder-type probe transitions are compared to predictions from the TheoReTS database [10].

The experimental setup is shown in Fig. 2. The pump was a singly-resonant optical parametric oscillator (OPO, Aculight, Argos 2400 SF, module C) with idler tunable in the 3.1-3.7 μm range with power of up to 1 W. The idler frequency was first tuned to the vicinity of the selected pump transition in the fundamental $\nu_3$ band using a wavemeter (Burleigh, WA-1500-NIR-89, ±0.0001 nm resolution, ±2×10$^{-7}$ absolute relative accuracy) that monitored the pump and signal frequencies. Then the idler frequency was stabilized to the center of the addressed transition using a $CH_4$ reference cell and a frequency modulated Lamb-dip lock (see Supplementary Material, ref. [23], for details of the stabilization scheme). This lock provided long-term frequency stability of the idler of ±0.8 MHz, limited by the width of the error signal.

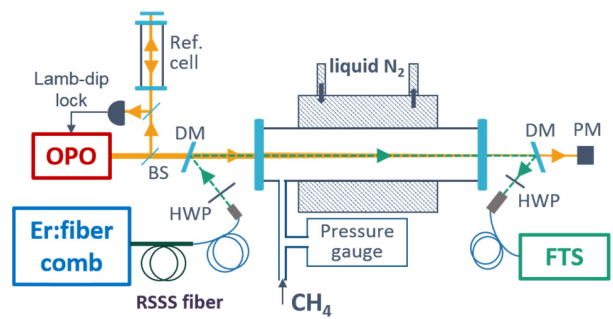

FIG. 2. (Color online) Experimental setup. OPO: optical parametric oscillator, BS: beam splitter, DM: dichroic mirror, PM: power meter, HWP: half-wave plate, RSSS fiber: Raman soliton self-frequency shift microstructured silica fiber, FTS: Fourier transform spectrometer.

The probe was an amplified Er:fiber frequency comb (Menlo Systems, FC1500-250-WG) with 250 MHz repetition rate and 400 mW output power around 1.55 μm. The comb spectrum was shifted to cover 200 cm$^{-1}$ around 1.67 μm (6000 cm$^{-1}$) with 20 mW of power using a custom-made polarization-maintaining Raman soliton self-frequency shift microstructured silica fiber [24]. The comb was RF-stabilized to a GPS-referenced Rb oscillator (Symmetricom, TSC 4410A), as described in ref. [23]. The carrier envelope offset frequency was fixed at 20 MHz, while the repetition rate, $f_{rep}$, was locked to a tunable direct digital synthesiser (DDS) that allowed stepping of the comb mode frequencies.

A sample of pure $CH_4$ was contained in an 80-cm-long single-pass cell with $CaF_2$ windows. The central 55 cm was liquid-nitrogen-cooled to increase the absorption signal. For the R(0) line in the $\nu_3$ fundamental band, the absorption coefficient at fixed pressure scales as T$^{-3}$ so the absorption signal is expected to be 56 times stronger at T = 77 K than at room temperature. The pump and probe beams were combined using a dichroic mirror (Layertec, pump mirror 105804) in front of the cell and separated with a similar mirror behind the cell. The OPO beam was collimated with a radius of 1 mm along the length of the cell, while the comb beam had a radius of 0.36 mm at the waist in the middle of the cell. The power of the transmitted pump beam was monitored using a power meter. The transmitted probe beam was coupled into a polarization-maintaining (PM) fiber and led to a home-built fast-scanning Fourier transform spectrometer (FTS) with auto-balanced detection [25]. A zero-order half-wave plate (Thorlabs, WPH05M-1550) in front of the cell was used to adjust the polarization of the probe to be parallel or perpendicular to the pump polarization. The second zero-order half-wave plate, positioned after the cell, was used to align the polarization of the comb light to the slow axis of the PM fiber. We note that the polarization of the probe was not purely linear when the half-wave plate was rotated because of the wide comb bandwidth.

The pressure in the cell was adjusted so that the transmission of the pump power on resonance was 20%, which was found to yield a maximum V-type probe signal. For the R(0) transition, this corresponded to 30 mTorr. The saturation intensity at 30 mTorr and 111 K (the actual sample temperature, see below) is 1.46 mW/mm$^2$ (assuming hard-sphere collisions), which means that the transition was strongly saturated, with the intensity at the entrance of the cell 180 times above saturation. The experimental conditions for all pumped lines are summarized in ref. [23].

The nominal resolution of the FTS was set to 250 MHz to enable precise measurements of comb mode intensities without the influence of the instrumental lineshape function [26]. One interferogram was acquired in 5.8 s and yielded sample point spacing of 250 MHz. The $f_{rep}$ was then tuned 125 times in steps of 2.78 Hz, which resulted in a ~2 MHz shift of the comb modes per step. These 125 spectra were interleaved and averaged (up to 16 times, see ref. [23]), to yield a noise equivalent absorption coefficient of 2 × 10$^{-6}$ cm$^{-1}$ after 3.2 h averaging. The measurement was then repeated with the probe polarization rotated by 90°. In total, DR spectra were recorded for the pump laser locked to 9 different transitions, viz. R(0), R(1), P(1), Q(1), P(2,E), P(2,F2), P(3,A2), P(3,F1), P(3,F2).

Figure 3 shows the probe spectrum recorded with the pump locked to the R(0) transition in the $\nu_3$ fundamental band and with perpendicular pump and probe polarizations. Most of the lines in the spectrum are Doppler-broadened transitions in the 2$\nu_3$ overtone band. A fit to the entire spectrum, described in detail in ref. [23], reveals that the sample temperature was 111(4) K, i.e. significantly above 77 K. We attribute this to inefficient cooling of the cell and contributions from the warmer gas in the uncooled part of the cell (temperature retrieved from a spectrum with pump beam off was found to be 100 K), as well as to the heating by the pump beam (predicted to yield 7 K temperature increase). The inset shows a zoom of the 2$\nu_3$ R(0) overtone transition, in which the sub-Doppler V-type probe transition is visible. This transition has a full width at half maximum of 12.9(2) MHz and a depth of 40% relative to the Doppler-broadened line. The predicted width and depth of this transition are 8 MHz and 50%, respectively, where the width is dominated by the power broadening of the pump transition. This indicates that the probe transition may be additionally broadened by the frequency jitter of the pump.

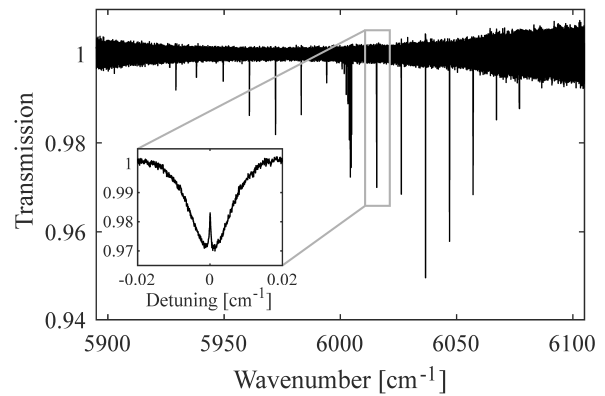

FIG. 3. (Color online) The 2$\nu_3$ overtone and DR spectrum of methane measured with the comb probe when the pump is locked to the fundamental R(0) transition and the relative pump/probe polarizations are perpendicular. The inset shows the V-type probe transition in the 2$\nu_3$ overtone R(0) line.

In total, for the 9 different pumped transitions we detected V-type resonances in 18 different overtone transitions, listed in Table 2 in ref. [23]. In Fig. 4, the center frequencies of five V-type resonances, obtained from a fit of a Lorentzian function, are compared to the center frequencies of the corresponding 2$\nu_3$ overtone lines measured in the Doppler-limited regime by Zolot *et al.* [27] using dual comb spectroscopy (DCS). The error bars in the figure are combinations of the uncertainties originating from the fit and the frequency stability of the pump (1.33 MHz, 4.4×10$^{-4}$ cm$^{-1}$, see ref. [23] for details), and the uncertainties reported for the DCS measurements, which are of the order of 0.2–0.6 MHz (0.6–2×10$^{-5}$ cm$^{-1}$). The agreement between the two measurements is very good, with the rms value of the discrepancies equal to 0.35 MHz, indicating that the frequency accuracy of DR measurements is below 1 MHz. We also note that the V-type transitions in the Q(1) and R(1) overtone lines were observed in three different measurement series, with pump locked to the R(1), Q(1) and P(1) lines in the fundamental band. These measurements were separated in time by 3 and 2 weeks, respectively. The center frequencies of the Q(1) and R(1) lines from the different measurements agree within their uncertainties, and their standard deviations are 0.1 MHz and 0.47 MHz, respectively. This indicates that the precision of the center frequencies determined using sub-Doppler DR spectroscopy is below 1 MHz even though the signal-to-noise ratio is at most 10 for the strongest lines. Moreover, the sub-Doppler V-type resonances allow determination of line center positions with the same accuracy/precision even in case of overlapping Doppler-broadened lines (e.g. the P(2) doublet, see ref. [23]), which is not possible with Doppler-limited spectroscopy.

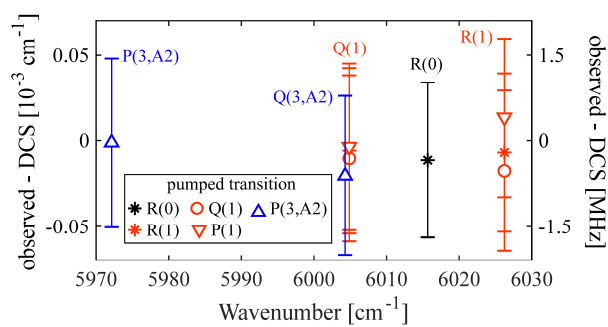

FIG. 4. (Color online) The difference between transition frequencies of the sub-Doppler V-type resonances in the R(0), R(1), Q(1), Q(3,A2) and P(3,A2) lines in the $2\nu_3$ overtone band (detected with pump locked to 5 different fundamental transitions, as marked in the legend), and the $2\nu_3$ overtone transitions from ref. [27]. The left axis is in wavenumbers, while the right axis is in MHz. Error bar – see text.

The ladder-type probe transitions were found in the spectra using a numerical procedure described in ref. [23]. Five such transitions were found when the pump was locked to the R(0) transition, and three of them are shown in Fig. 5 for perpendicular (black) and parallel (red) pump/probe polarizations. The center frequency, the area and the width of each transition was found by fitting a Lorentzian function to the data, together with a second-order polynomial to model the weak Doppler-broadened background caused by elastic collisions. The fits are shown by the thicker curves in Fig. 5, and the fitting routine and results for all DR transitions are summarized in section IV of ref. [23]. The uncertainty of the transition frequencies, calculated as a combination of the fit uncertainty and the stability of the pump, ranges from 1.36 to 3 MHz, with a mean of 1.7 MHz.

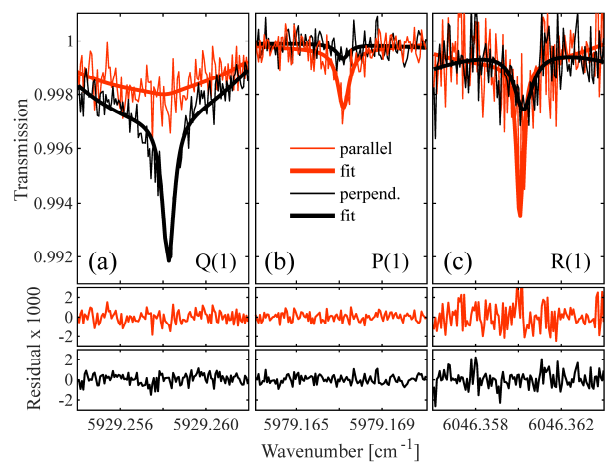

FIG. 5. (Color online) The (a) Q(1), (b) P(1) and (c) R(1) ladder-type probe transitions observed in the spectrum when the pump is locked to the fundamental R(0) line, recorded with perpendicular (black) and parallel (red) relative pump/probe polarizations. The thin curves show the data, the thick curves show the fits, and the residuals are shown in the lower panels.

The line assignment indicated in Fig. 5 is based on the ratios of transition intensities recorded with parallel and perpendicular pump/probe polarizations. The expected ratios for pump on the R(0) transition, calculated from the ratios of squared transition matrix elements summed over M quantum numbers, are 0 for a Q(1) line, infinity for a P(1) line and 1.33 for a R(1) line [28], in good agreement with observations. The assignment is confirmed by comparison with theoretical predictions from the TheoReTS database [10], which contains rotationally resolved hot methane line lists obtained from *ab initio* calculations (with empirical corrections to band centers applied when available) with a 0.1-1 $cm^{-1}$ estimated accuracy [29]. The differences between the experimentally determined center frequencies of the ladder-type probe transitions and predictions from TheoReTS are shown in Fig. 6 (the experimental error bars, which are negligible on this scale, are not shown). Remarkably, all detected lines are within 1 $cm^{-1}$ from predictions, while the frequency difference between predicted lines sharing the lower level is a few tens of wavenumbers. In total, 36 ladder-type probe transitions have been identified and assigned, of which only 10 were previously observed, though with much lower frequency accuracy [30-32]. The experimental and predicted transition frequencies of all detected lines are listed in ref. [23] together with the line intensities from TheoReTS, and the final state term values, calculated as the sum of pump transition frequencies and ground state upper-state term value from refs [33,34] and the experimental probe transition frequency. To further confirm the line assignment, the experimental intensity ratios of the probe transitions in the ladder- and V-type excitations are calculated in ref. [23] and compared to predictions based on the Einstein A coefficients from the TheoReTS and HITRAN [35] databases.

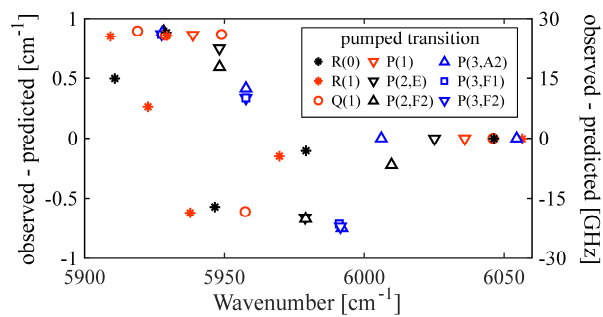

FIG. 6. (Color online) The difference between center frequencies of the sub-Doppler ladder-type probe transitions (detected with the pump tuned to 9 different fundamental transitions, as marked in the legend), and the center frequencies predicted in the TheoReTS database. The left axis is in wavenumbers, while the right axis is in GHz.

In conclusion, we report the first measurement of sub-Doppler molecular response using a frequency comb and detect 26 previously unobserved ladder-type double-resonance transitions in methane with 1.7 MHz center frequency accuracy. All transitions are assigned with the help of the TheoReTS line list and confirmed by line intensity ratios for lines with low J numbers. The frequency accuracy is currently limited mainly by the stability of the pump frequency (1.33 MHz), which can be improved by optimizing the Lamb dip lock, or replacing it with a direct lock of the pump frequency to the comb. The signal-to-noise ratio of the strongest ladder-type probe transitions is currently limited to around 10 by the short interaction length. The absorption sensitivity can be increased by implementing an enhancement cavity for the comb probe. These improvements will allow detection of a larger number of weaker probe transitions with higher frequency accuracy and larger precision on the line shape, and thus confirming the assignment for higher J states. The assigned levels in the 9000 $cm^{-1}$ energy region can serve as stepping stones to access even higher levels in the 15000 $cm^{-1}$ range using a 1 μm pump.

Double-resonance spectroscopy using two monochromatic lasers is a well-established method, but the frequency range that can be probed with it is restricted by the limited tunability of the narrow-linewidth lasers. Using a frequency comb probe, double-resonance transitions can be detected with tens of thousands of narrow comb teeth simultaneously. The technique thus opens up for accurate assignment of many highly excited energy levels and allows verification of theoretical predictions of spectral transition frequencies and intensities. Such verification is needed to assess the accuracy of predicted high-temperature spectra for modeling high-temperature environments such as exoplanets. The technique, with different pump/probe configurations, is broadly applicable to the assignment of spectra of many molecules, for which spectral perturbations and/or congestion prevent assignment by traditional methods alone.

The authors thank Michael Rey for providing additional information from the TheoReTS database, Hiroyuki Sasada for providing the ground state upper state term values, and Jin Guo and Chuang Lu for help with setting up the OPO. Aleksandra Foltynowicz acknowledges the Swedish Research Council (2016-03593) and the Knut and Alice Wallenberg Foundation (KAW 2015.0159). Grzegorz Soboń acknowledges the Foundation for Polish Science (POIR.04.04.00-00-434D/17-00). Ove Axner acknowledges the Swedish Research Council (2015-04374) and the Kempe Foundations (JCK 1317.1).

*Supplemental Material for*
# Sub-Doppler Double-Resonance Spectroscopy of Methane Using a Frequency Comb Probe

Aleksandra Foltynowicz[1], Lucile Rutkowski[2], Isak Silander[1], Alexandra C. Johansson[1], Vinicius Silva de Oliveira[1], Ove Axner[1], Grzegorz Soboń[3], Tadeusz Martynkien[4], Paweł Mergo[5], Kevin K. Lehmann[6]

[1]*Department of Physics, Umeå University, 901 87 Umeå, Sweden*
[2]*Univ Rennes, CNRS, IPR (Institut de Physique de Rennes)-UMR 6251, F-35000 Rennes, France*
[3]*Laser & Fiber Electronics Group, Faculty of Electronics, Wrocław University of Science and Technology, 50-370 Wrocław, Poland*
[4]*Faculty of Fundamental Problems of Technology, Wrocław University of Science and Technology, 50-370 Wrocław, Poland*
[5]*Laboratory of Optical Fiber Technology, Maria Curie-Sklodowska University, pl. M. Curie-Sklodowskiej 3, 20-031 Lublin, Poland*
[6]*Departments of Chemistry & Physics, University of Virginia, Charlottesville, VA 22904, USA*
e-mail address: aleksandra.foltynowicz@umu.se; kl6c@virginia.edu

## I. EXPERIMENTAL CONDITIONS

Table 1 summarizes the experimental conditions of the measurements of all spectra for the 9 different pumped transitions in the $v_3$ fundamental band, listed in column 1. Column 2 shows the $CH_4$ pressure in the cell, while columns 3 and 4 show the corresponding linear absorption coefficient, α, and the saturation intensity calculated at a temperature of 111 K (see section III.2 below). Column 5 shows the pump power measured in front of and behind the cell (with pump locked to the transition), and column 6 shows the number of spectra averaged. For some spectra the number of averages is different for parallel, P, and crossed, X, relative pump/probe polarizations, since spectra for which the OPO unlocked from the Lamb dip were removed.

Table 1: Experimental conditions for the different measurement series. X and P – perpendicular and parallel relative pump/probe polarizations.

| Pumped transition | Pressure [mTorr] | Linear α @ 111 K [m$^{-1}$] | Saturation intensity [mW/mm$^2$] | Pump power input / output [mW] | #averages |
|---|---|---|---|---|---|
| R(0) | 30 | 17.1 | 1.46 | 850 / 150 | 16 X / 11 P |
| R(1) | 30 | 15.1 | 4.38 (M=0), 2.19 (M=±1) | 890 / 170 | 16 |
| Q(1) | 35 | 10.3 | 3.53 (M=±1) | 890 / 170 | 16 |
| P(1) | 60 | 5.8 | 17.5 (M=0) | 850 / 152 X<br>867 / 175 P | 17 X / 5 P |
| P(2,E) | 58 | 8.5 | 13.6 (M=0), 18.2 (M=±1) | 1130 / 208 | 15 X / 11 P |
| P(2,F2) | 40 | 8.8 | 6.49 (M=0), 8.65 (M=±1) | 940 / 194 | 10 X / 16 P |
| P(3,A2) | 33 | 13.2 | 4.12 (M=0), 4.84 (M=±1), 12.4 (M=±2) | 850 / 126 | 16 X / 8 P |
| P(3,F1) | 33 | 7.9 | 4.12 (M=0), 4.84 (M=±1), 12.4 (M=±2) | 850 / 156 | 16 X / 12 P |
| P(3,F2) | 33 | 7.9 | 4.12 (M=0), 4.84 (M=±1), 12.4 (M=±2) | 700 / 140 | 16 X / 9 P |

## II. PUMP AND PROBE FREQUENCY STABILIZATION

### 1. OPO stabilization

The idler frequency of the optical parametric oscillator (OPO, Aculight, Argos 2400 SF, module C) was stabilized to the center of a chosen CH$_4$ transition in the fundamental $\nu_3$ band using a frequency modulated (FM) Lamb dip lock, as shown in Fig. S 1(a). Part of the beam (50 mW) was directed to a 16.5 cm long reference cell filled with a few tens of mTorr of pure methane. The light was back reflected, overlapped with the incoming beam, and directed on a high-bandwidth HgCdTd photodiode (PD, VIGO systems, PVI-4TE-8-1x1) using a beamsplitter. The phase of the pump light was modulated at 60 MHz ($f_m$) with a modulation index of 0.3 using an electro-optic modulator (EOM, Photline, NIR-MPX-LN-05) inserted between the seed (NKT Photonics, Koheras Adjustik Y-10) and the amplifier (IPG Photonics, YAR-10) of the cw Yb:fiber laser. The phase modulation of the pump light is transferred to the idler [36]. The detector signal was synchronously demodulated at $f_m$ to yield an FM error signal, shown in Fig. S 2. The detection phase was adjusted using a phase shifter to yield the flattest possible baseline (originating from the Doppler-broadened background). The error signal was fed to a proportional-integral controller (Newport, LB1005) and the correction was applied to a PZT that controls the length of the fiber in the Yb:fiber seed laser.

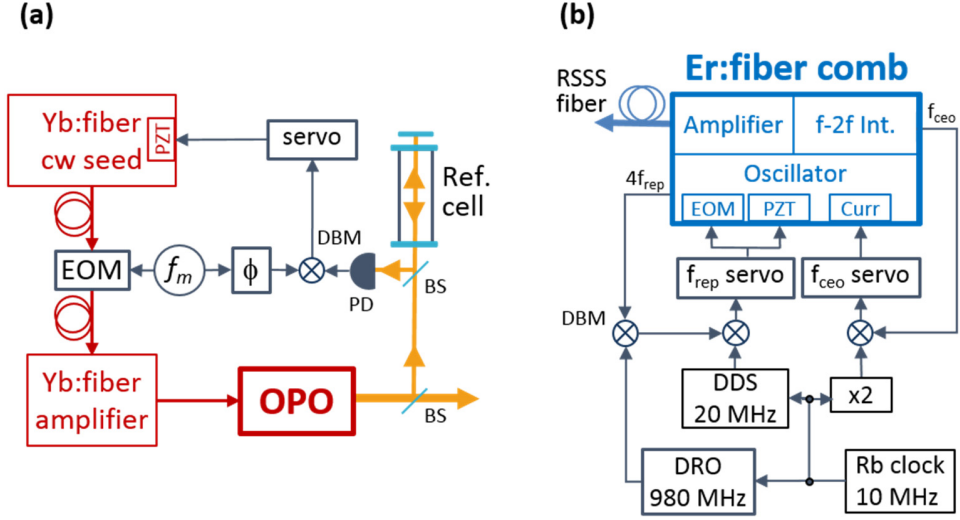

Fig. S 1. (Color online) (a) Stabilization of the OPO idler frequency. EOM: electro-optic modulator, PZT: piezoelectric transducer, $f_m$: function generator, $\phi$: phase shifter, DBM: double balanced mixer, PD: photodetector, BS: beamsplitter. (b) Stabilization of the frequency comb. RSSS fiber: Raman soliton self-frequency shift microstructured silica fiber, EOM: electro-optic modulator, PZT: piezoelectric transducer, Curr: pump diode current, f-2f Int.: f-2f interferometer, DBM: double balanced mixer, DDS: direct digital synthesizer, DRO: dielectric resonator oscillator.

The width of the central peak of the error signal was found by fitting a model based on sub-Doppler FM spectroscopy formalism [37] to the open loop error signal, shown in Fig. S 2. The error signal as a function of frequency detuning from the center, $\Delta\nu$, is given by

$$S(\Delta\nu, f_m, \phi, \Gamma, b) = I_0 \left\{ \left[ \chi^{disp}(\Delta\nu - f_m/2, b\Gamma) - 2\chi^{disp}(\Delta\nu, \Gamma) + \chi^{disp}(\Delta\nu + f_m/2, b\Gamma) \right] \cos\phi \right. \\ \left. + \left[ \chi^{abs}(\Delta\nu - f_m/2, b\Gamma) - \chi^{abs}(\Delta\nu + f_m/2, b\Gamma) \right] \sin\phi \right\}. \quad (1)$$

where $\chi^{abs}(\Delta\nu, \Gamma)$ is the area-normalized Lorentzian function with half width at half maximum (HWHM) width $\Gamma$, and $\chi^{disp}(\Delta\nu, \Gamma)$ is its dispersion counterpart. The error signal consists of three components, where the side peaks are separated from the central peak by half of the modulation frequency, $\pm f_m/2$ (see Fig. 7.6 in ref. [38]). The central peak originates mostly from a carrier-carrier interaction, i.e. the Bennett hole burned by the carrier field and detected by the counter-propagating carrier. At a detuning equal to half the modulation frequency, the sub-Doppler absorption signal comes from the carrier-sideband interaction, i.e., the Bennett hole burned by the carrier and detected by the counter-propagating FM sideband, while the dispersion signal has an additional contribution from the Bennett hole burned by the sideband and detected with the carrier. This implies that the degrees of saturation are different for the center and side peaks, which leads to different power broadening. This in turn causes the width of the central and side peaks to be different. To reflect that, the width of the side peaks is multiplied by a scaling factor $b$. Furthermore, $I_0$ is an amplitude and $\phi$ is the detection phase.

Figure S 2 shows the open loop error signal recorded for the R(0) line at 70 mTorr with 50 mW of incident power (black), together with a fit of Eq. (1) (red). A second order polynomial was added to the fit to model the Doppler-broadened background (the HWHM of the Doppler-broadened lines in the fundamental band at room temperature is 140 MHz, which results in a non-negligible FM signal when the modulation frequency is 60 MHz). The frequency scale was calibrated using the zero crossings of the three peaks as markers separated by 30 MHz. The fitting parameters were $I_0$, $\phi$, $\Gamma$ and $b$, and the fit returned $\Gamma$ = 2 MHz, $b$ = 0.67 and $\phi$ = 0.42. The sharp features in the residuum (lower panel) at detunings of -30 MHz, 0 MHz, and 30 MHz originate from inaccuracies in frequency scale calibration, while the slowly varying structure is the remaining Doppler-broadened background and etalon signals.

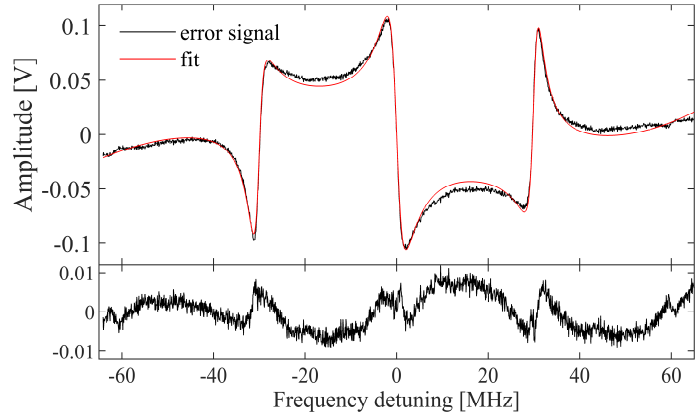

Fig. S 2. (Color online) Frequency modulated Lamb dip error signal (top panel, black) with fit (red) and residuum (lower panel).

We found that the offset of the error signal drifted over time because of residual amplitude modulation in the OPO output, and that the idler unlocked when this offset was equal to 40% of the peak value. This implies that the idler frequency was stable to within $\pm 0.4\,\Gamma = \pm 0.8$ MHz. This translates to an uncertainty in the center frequency of the double-resonance transitions equal to $\pm 1.6$ MHz because the ratio of the probe to pump frequencies is equal to 2, and thus the Doppler shifts produced by an error in the Lamb dip lock are twice larger for the probe than for the pump. In our uncertainty analysis we assumed conservatively that the maximum drift of the center frequency of the probe excitation in the double-resonance transitions caused by the pump frequency drift was 2 MHz, and that the 1σ uncertainty was equal to 2/3 of that, i.e. 1.33 MHz.

## 2. Comb stabilization

The Er:fiber frequency comb (Menlo Systems, FC1500-250-WG) was RF-stabilized to a GPS-referenced Rb oscillator (Symmetricom, TSC 4410A) as shown in Fig. S 1(b). The comb carrier envelope offset frequency, $f_{ceo}$, was measured using an f-2f interferometer and stabilized at 20 MHz (twice the Rb oscillator frequency) by feeding back to the pump diode current of the oscillator. The 4th harmonic of the repetition rate, $f_{rep}$, was mixed with a stable 980 MHz signal from a dielectric resonator oscillator (DRO), referenced to the Rb oscillator. The resulting signal around 20 MHz was locked to a tunable direct digital synthesiser (DDS) that used the Rb oscillator as a clock. Feedback for $f_{rep}$ stabilization was sent to a PZT and EOM inside the oscillator cavity. The DDS frequency was stepped by 2.78 Hz to shift the comb repetition rate to produce spectra for interleaving.

## III. DATA TREATMENT AND FITTING

### 1. Sub-nominal procedure optimization, baseline removal and interleaving

The optical path difference in the Fourier transform spectrometer was calibrated using a stabilized HeNe laser. The sub-nominal resolution approach requires iterative optimization of the wavelength of the reference cw laser, as described in ref. [26]. The optimum wavelength was found by fitting a Gaussian function to the Doppler-broadened $2\nu_3$ overtone $CH_4$ transition in the interleaved spectrum corresponding to the pumped transition in the $\nu_3$ fundamental band. The optimum wavelength is the one for which both the residuum of the fit and the width of the line are minimized. We note that the position and line shape of the sub-Doppler probe transitions is unaffected by any remaining uncertainty of the cw reference laser wavelength, see Figs 4-6 in ref. [26].

Spectra taken with the same $f_{rep}$ value were averaged. Next, the baseline was removed using cepstral analysis similar to that described in ref. [39]. This method relies on analysing the spectrum in the time rather than the frequency domain, where the contributions of the baseline and the molecular signal are easier to separate. In particular, the contribution of a slowly varying baseline is present mostly at the shorter times, while the free induction decay of the molecular signal is present at longer times as well. To remove the baseline, first, the Doppler-broadened transmission spectrum of the $2\nu_3$ band of $CH_4$ was calculated for the pertinent experimental conditions (temperature, pressure and cell length) using the line parameters from the HITRAN database [35]. Next, the inverse fast Fourier transform (iFFT) of this model spectrum was fit to the iFFT of the measured spectrum with amplitude (i.e. concentration) as the only fitting parameter. A step weighing function was applied to both the model and the data to remove the first N points from the fit, where the baseline contribution is strongest. Afterwards, a residual of the fit was calculated over all data points and the first N points of the residual were transformed back to the frequency domain. This yielded a smoothed baseline that was used to normalize the transmission spectra. The same procedure was repeated for the 125 individual spectra and the resulting spectra were interleaved.

### 2. Doppler-broadened fits

The temperature of the sample was estimated by fitting a model to the central part of the Doppler-broadened $2\nu_3$ overtone band [i.e. lines P(4) through R(4), see Fig. S 3]. In the model, the transition parameters were taken from the HITRAN database [35], and the optical path (55 cm) and pressure (see Table 1) were fixed. The only fitting parameter was temperature, which affects the Doppler width and the relative intensities of all lines. The temperature dependence of the line strength was modelled using Eq. A11 in ref. [40]. A fit to a Doppler-broadened spectrum recorded with the pump laser off yielded a temperature of 100 K. This is higher than 77 K and we attribute this to inefficient cooling by liquid nitrogen and the warmer gas in the 31% of the path length outside of the dewar. The fits to spectra taken with pump locked to the different transitions returned a temperature of 111(4) K, where the uncertainty is the standard deviation for all data sets. The increase of the temperature with the pump on is in good agreement with numerical simulations of heat transport, which yield 7 K radial temperature increase for the pump locked to the R(0) line.

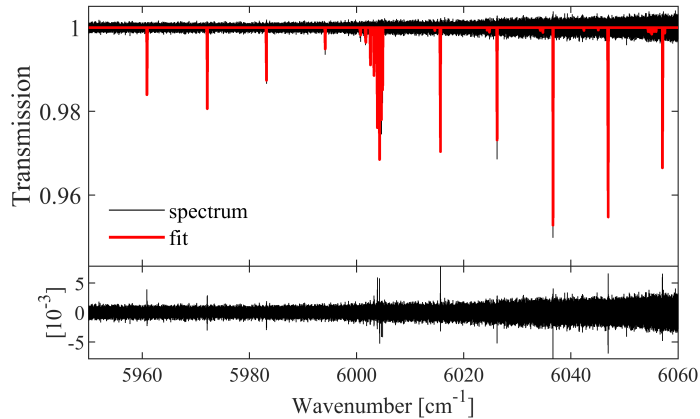

Fig. S 3. (Color online) Part of the Doppler-broadened spectrum (black) with pump locked to the R(0) fundamental transition together with a fit of a model (red) used to extract the temperature of the sample. The residuum is shown in the lower panel.

### 3. Sub-Doppler fits

In order to find the ladder-type transitions, the interleaved spectrum was first divided by a model of the Doppler-broadened $2\nu_3$ overtone spectrum, based on the HITRAN parameters [35] and the pertinent experimental conditions, to remove it from the spectrum. Next, the spectrum was convolved with a 20 MHz wide (HWHM) Lorentzian dispersive function to reduce the influence of the noise and the remainder of the Doppler-broadened lines. Next, the first derivative of the spectrum was calculated. Since the noise level across the spectrum was varying, the spectrum was normalized by the moving mean of the square of the spectrum calculated over a window of 10000 points (i.e. 20 GHz, or 0.667 cm$^{-1}$). Finally, a peak finding routine (Matlab 'findpeaks' function) was applied with a threshold set to detect peaks with SNR>1. All detected peaks were inspected visually to reject false detections.

To retrieve the center frequency, width and area of each double-resonance transition a Lorentzian function was fit to the unmodified interleaved spectrum. The fitting range was ±0.004 cm$^{-1}$ around each transition. First, the central part of this range (±0.001 cm$^{-1}$) was masked, and a 2$^{nd}$ order polynomial was fit to the background and subtracted from the data to remove the baseline. After that, the Lorentzian function was fit with center frequency, width and amplitude as fitting parameters using a modified Levenberg-Marquardt algorithm. The data and the fits to probe transitions detected with the pump locked to the R(0) transition in the fundamental band are shown in Fig. S 4 for the V-type (a) and the ladder-type (b)-(f) excitations. The uncertainties (standard errors) of the fit parameters were calculated as the square root of the diagonal covariance matrix element. The quality factor (QF) of each fit was calculated as the ratio of the fit Lorentzian amplitude and the rms of the residual. The fit parameters for all lines are summarized in section IV below.

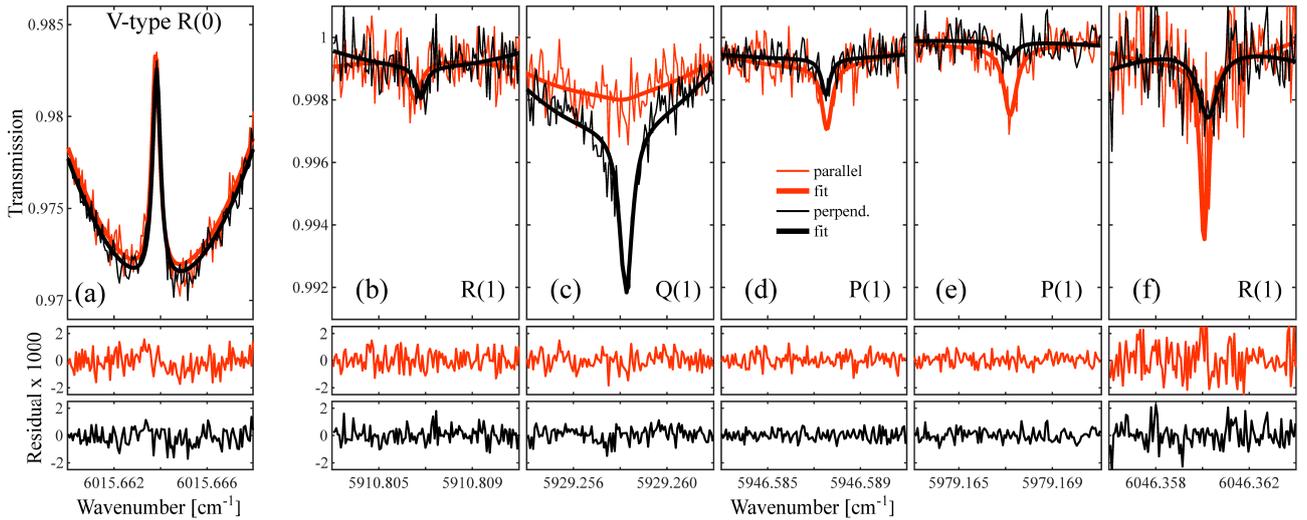

Fig. S 4. (Color online) The (a) V-type and (b)-(f) ladder-type double resonances detected with pump laser locked to the R(0) fundamental transition for parallel (red) and perpendicular (black) relative pump/probe polarizations. The thin curves in the upper panels show the data, and the thick lines show the fits of the Lorentzian model and 2$^{nd}$ order polynomial baseline. The two lower panels show residuals of the fits.

## IV. LINE PARAMETERS

### 1. Center frequencies

The wavenumbers of the double-resonance transitions were calculated as a weighted mean of the wavenumbers found from the fit to the probe transitions measured using parallel and perpendicular relative pump/probe polarizations. The inverse of the square of the fit uncertainty (i.e. the variance) was used as a weight. The weight was put to zero for all lines for which the QF was less than 2 [e.g. the red curve in Fig. S 4(c) or the black curve in Fig. S 4(e)]. The standard deviation of the weighted mean was calculated using the assumption that the inverse of the variance of the weighted mean is equal to the sum of the inverse of the variances of the two values. The uncertainty originating from the drift of the pump center frequency (1.33 MHz, see section II.1 above) was added in squares to the resulting uncertainty.

The wavenumbers of the V-type probe transitions are summarized in column 3 of Table 2. Column 1 indicates the pumped transition in the $v_3$ fundamental band and column 2 indicates the probed transition in the $2v_3$ overtone band.

The wavenumbers of the observed ladder-type probe transitions are summarized in column 3 of Table 3. Column 1 indicates the pumped transition in the $v_3$ fundamental band and its upper state term value (corresponding to the lower state term value of the probe transitions) taken from refs [33,34]. Column 2 shows the assignment of the probe transitions and column 4 shows the corresponding transition wavenumbers obtained from the TheoReTS database [10]. They were identified by searching the database for transitions with a lower state energy corresponding to the upper level of the pumped transitions within a ±0.05 cm$^{-1}$ window. When the lower levels listed in TheoReTS were not equal to the HITRAN energy levels, the energy difference $E_{HITRAN}$ - $E_{TheoReTS}$ was subtracted from the TheoReTS transition wavenumbers to correct for this discrepancy. This procedure allowed identifying 3-10 TheoReTS lines within the measured spectral range for each pumped transition. The strongest lines matched the detected probe transitions within 1 cm$^{-1}$. The assignment of the measured transitions was then deduced from the upper and lower J values of the corresponding predicted transitions and confirmed by the experimental intensity ratios in spectra recorded with parallel and perpendicular relative pump/probe polarizations for low J numbers. Column 5 shows the difference between the observed and the predicted transition wavenumbers, and column 6 shows the line intensities from the TheoReTS database calculated at 2000 K from *ab initio* Einstein A coefficients. Column 7 shows the final state term value, calculated as the sum of the lower energy level of the probe transition (column 1) and the experimental probe transition frequency (column 3). The last column shows the term values of the final states reached in the measurements of De Martino *et al*. [30-32].

Table 2: Wavenumbers of V-type probe transitions, corresponding to transition frequencies in the $2v_3$ overtone band.

| Pumped transition | Probe transition | Transition wavenumber [cm$^{-1}$] |
|---|---|---|
| R(0) | R(0) | 6015.66382(4) |
| R(1) | R(1) | 6026.22685(5) |
| | Q(1) | 6004.86265(5) |
| | P(1) | 5994.14368(5) |
| Q(1) | R(1) | 6026.22684(5) |
| | Q(1) | 6004.86264(5) |
| | P(1) | 5994.14370(5) |
| P(1) | R(1) | 6026.22687(5) |
| | Q(1) | 6004.86265(5) |
| | P(1) | 5994.14371(5) |
| P(2,E) | R(2,E) | 6036.65385(5) |
| | P(2,E) | 5983.18341(5) |
| P(2,F2) | R(2,F2) | 6036.65764(5) |
| | Q(2,F2) | 6004.64356(5) |
| | P(2,F2) | 5983.19366(5) |
| P(3,A2) | R(3,A2) | 6046.96358(5) |
| | Q(3,A2) | 6004.29224(5) |
| | P(3,A2) | 5972.13391(5) |
| P(3,F1) | R(3,F1) | 6046.94204(5) |
| | Q(3,F1) | 6004.32864(5) |
| | P(3,F1) | 5972.09529(5) |
| P(3,F2) | R(3,F2) | 6046.95168(5) |
| | Q(3,F2) | 6004.31281(5) |
| | P(3,F2) | 5972.11235(5) |

Table 3: Experimental and predicted wavenumbers of ladder-type probe transitions, together with predicted line intensities, the experimental final state values.

| Pumped transition and its upper state term value [cm$^{-1}$] | Probe transition | Transition wavenumber [cm$^{-1}$] | Transition wavenumber TheoReTS [cm$^{-1}$] | Obs-Calc [cm$^{-1}$] | Linestrength at 2000 K TheoReTS [10$^{-25}$ cm/mol] | Final state term value [cm$^{-1}$] | Final state term value ref. [30-32] [cm$^{-1}$] |
|---|---|---|---|---|---|---|---|
| R(0) 3028.752260 | R(1) | 6046.36008(5) | 6046.36012 | -0.000037 | 1.732 | 9075.11234(5) | |
| | P(1) | 5979.16720(5) | 5979.27374 | -0.11 | 0.6111 | 9007.91946(5) | |
| | P(1) | 5946.58753(5) | 5947.15944 | -0.57 | 0.8031 | 8975.33979(5) | |
| | Q(1) | 5929.25825(5) | 5928.37576 | 0.88 | 5.367 | 8958.01051(5) | 8957.96 |
| | R(1) | 5910.8068(1) | 5910.30937 | 0.5 | 1.127 | 8939.5591(1) | 8939.56 |
| R(1) 3048.980142 | R(2) | 6056.48376(5) | 6056.48569 | -0.0019 | 1.385 | 9105.46390(5) | |
| | P(2) | 5969.66090(6) | 5969.8125 | -0.15 | 0.5347 | 9018.64104(6) | |
| | P(2) | 5937.7677(1) | 5938.38891 | -0.62 | 0.6879 | 8986.7478(1) | |
| | Q(2) | 5929.28494(5) | 5928.42915 | 0.86 | 3.926 | 8978.26508(5) | |
| | R(2) | 5922.67986(6) | 5922.41619 | 0.26 | 0.7666 | 8971.66000(6) | 8971.67 |
| | P(2) | 5909.24115(9) | 5908.3918 | 0.85 | 1.179 | 8958.22129(9) | 8958.68 |
| Q(1) 3029.306115 | R(1) | 6046.06828(6) | 6046.0679 | 0.00038 | 0.8117 | 9075.37440(6) | |
| | Q(1) | 5957.44166(5) | 5958.05332 | -0.61 | 0.5308 | 8986.74778(5) | |
| | R(1) | 5948.95899(5) | 5948.09357 | 0.87 | 1.335 | 8978.26511(5) | |
| | Q(1) | 5928.91510(5) | 5928.05621 | 0.86 | 0.8685 | 8958.22122(5) | |
| | P(1) | 5918.93759(5) | 5918.04234 | 0.9 | 1.077 | 8948.24371(5) | 8948.24 |
| P(1) 3019.493054 | R(0) | 6035.94676(5) | 6035.94784 | -0.0011 | 0.3874 | 9055.43981(5) | |
| | R(0) | 5938.72822(5) | 5937.86487 | 0.86 | 1.118 | 8958.22127(5) | |
| P(2,E) 3030.502542 | Q(1) | 6025.12811(6) | 6025.12942 | -0.0013 | 0.5527 | 9055.63065(6) | |
| | R(1) | 5978.87023(6) | 5979.53197 | -0.66 | 0.4273 | 9009.37277(6) | |
| | R(1) | 5948.19655(5) | 5947.44947 | 0.75 | 1.672 | 8978.69909(5) | 8978.68 |
| P(2,F2) 3030.436420 | R(1) | 6009.74667(5) | 6009.9696 | -0.22 | 0.3994 | 9040.18309(5) | |
| | R(1) | 5979.04297(5) | 5979.71487 | -0.67 | 0.6371 | 9009.47939(5) | |
| | R(1) | 5948.26760(5) | 5947.67461 | 0.59 | 2.043 | 8978.70402(5) | 8978.25 |
| | Q(1) | 5928.30687(6) | 5927.41421 | 0.89 | 0.7965 | 8958.74329(6) | |
| P(3,A2) 3051.673466 | R(2) | 6054.50217(6) | 6054.5022 | -0.000027 | 1.626 | 9106.17564(6) | |
| | P(2) | 6006.10212(6) | 6006.1016 | 0.00052 | 1.765 | 9057.77559(6) | |
| | R(2) | 5992.03737(5) | 5992.78685 | -0.75 | 1.801 | 9043.71084(5) | |
| | R(2) | 5957.70612(5) | 5957.29062 | 0.42 | 5.632 | 9009.37959(5) | |
| | Q(2) | 5927.864(2) | 5926.98221 | 0.88 | 2.818 | 8979.538(2) | 8979.52 |
| P(3,F1) 3051.909263 | R(2) | 5991.06680(6) | 5991.78249 | -0.72 | 1.088 | 9042.97593(6) | |
| | R(2) | 5957.67296(5) | 5957.336 | 0.34 | 3.136 | 9009.58209(5) | 9009.45 |
| | Q(2) | 5927.52678(5) | 5926.66173 | 0.87 | 1.803 | 8979.43591(5) | |
| P(3,F2) 3051.809388 | R(2) | 5991.50002(6) | 5992.23818 | -0.74 | 1.082 | 9043.30941(6) | |
| | R(2) | 5957.69395(5) | 5957.35843 | 0.34 | 3.122 | 9009.50334(5) | 9009.45 |
| | Q(2) | 5927.67308(5) | 5926.80471 | 0.87 | 1.749 | 8979.48247(5) | |

## 2. Line intensities

The absolute line intensities of ladder-type probe transitions are difficult to obtain from the measurements because the exact population in the lower level of the transitions is not known. However, if one assumes equal collision relaxation rates for the upper and lower states of the pump transition, the steady state population in the pumped state should match the depletion of the ground state, which can be estimated from the fractional absorption dip in the V-type transitions. Therefore, we normalize the areas of the ladder-type probe transitions to the area of the V-type probe transition in the Doppler-broadened line corresponding to the pumped transition, which removes the dependence on the population of the intermediate state. We then compare these ratios to theoretical predictions based on HITRAN and TheoReTS.

The area of each double-resonance probe transition was calculated as the product of the fit peak value and width (HWHM) multiplied by $\pi$. The weighed mean of the areas measured with parallel and perpendicular relative pump/probe polarizations was calculated, where the weight was 1 for parallel polarization and 2 for perpendicular. Lastly, the weighted area of each ladder-type probe transition was normalized to the weighted area of the corresponding V-type probe transition.

The predicted intensity ratios were found by calculating the factors $\Phi = A g_j / v_{ij}^2$, where $A$ is the spontaneous emission rate of the transition, $g_j$ is the upper state degeneracy, and $v_{ij}$ is the transition frequency. HITRAN parameters were used for the $2v_3$ band transitions (V-type) and TheoReTS predictions were used for the ladder-type probe transitions.

Figure S 5 shows the logarithms (base 10) of the experimental (red markers) and predicted (black markers) intensity ratios. The different panels correspond to different pumped transitions, as marked in the panel title. Table 4 lists the mean and standard deviation of the relative discrepancies between the logarithms of the experimental and predicted intensity ratios. The reasonably good agreement further supports the assignments as the theoretical intensities span a range of 0.38-5.6 × $10^{-25}$ cm/molecule.

Table 4: Mean and standard deviation (in parentheses) of the difference between the logarithms (base 10) of the experimental and predicted intensity ratios shown in Fig. S 5.

| Pumped transition | Mean of log of ratios |
|---|---|
| R(0) | -0.45(19) |
| R(1) | -0.204(95) |
| Q(1) | -0.112(71) |
| P(1) | 0.340(21) |
| P(2,E) | 0.14(14) |
| P(2,F2) | 0.360(33) |
| P(3,A2) | 0.16(33) |
| P(3,F1) | 0.138(15) |
| P(3,F2) | 0.036(53) |

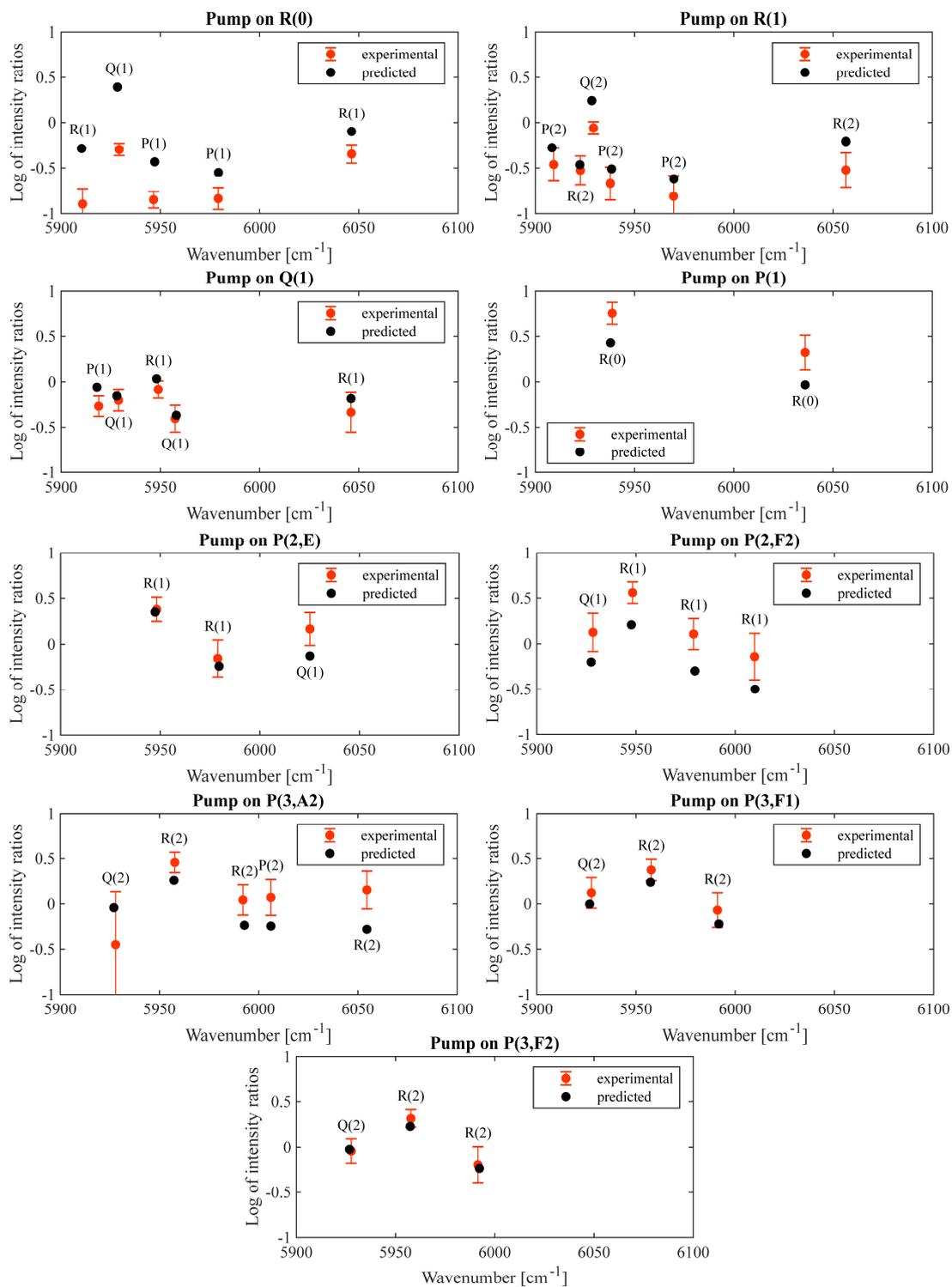

Fig. S 5. Logarithm (base 10) of the experimental (red markers) and predicted (black markers) intensity ratios of the probe transitions in the ladder-type and V-type excitations. See text for details.